\documentclass[twocolumn,showpacs,preprintnumbers,amsmath,amssymb]{revtex4}

\usepackage{graphicx}
\usepackage{dcolumn}
\usepackage{bm}
\usepackage[dvipdfm]{xcolor}

\begin{document}



\title{Atomic spatial coherence with spontaneous emission in a strong coupling cavity}

\author{Zhen Fang}
\affiliation{School of Electronics Engineering $\&$ Computer
Science, Peking University, Beijing 100871, China}
\author{Rui Guo}
\affiliation{School of Electronics Engineering $\&$ Computer
Science, Peking University, Beijing  100871, China}
\author{Xiaoji Zhou}\thanks{Electronic address: xjzhou@pku.edu.cn }
\affiliation{School of Electronics Engineering $\&$ Computer
Science, Peking University, Beijing  100871, China}
\author{Xuzong Chen}
\affiliation{School of Electronics Engineering $\&$ Computer
Science, Peking University, Beijing 100871, China}

\date{\today}

\begin{abstract}

The role of spontaneous emission in the interaction between a
two-level atom and a pumped micro-cavity in the strong coupling
regime is discussed in this paper. Especially, using a quantum
Monte-Carlo simulation, we investigate atomic spatial coherence. It
is found that atomic spontaneous emission destroys the coherence
between neighboring lattice sites, while the cavity decay does not.
Furthermore, our computation of the spatial coherence function shows
that the in-site locality is little affected by the cavity decay,
but greatly depends on the cavity pump amplitude.

\end{abstract}

.

\maketitle

\textbf{Introduction--} The combination of cold atom physics and
cavity quantum electrodynamics (cavity QED) have made possible the
investigation of the coherence property of matter wave in periodical
potential~\cite{nat404, nat428, prl95,pra64, jpb38, prlZippilli, ol21, np3, prl98,
oc273, epjd46}. A tunable optical lattice can be generated by
pumping a single mode micro-cavity with a far detuned laser, and
strong coupling between the atom(s) and the cavity field can be
reached. In this regime the recoil by scattering photons can be very
important~\cite{prl98-2, pra80}. Even a single photon may transfer
significant momentum to the atom(s) and reversely the atomic
distribution also strongly affects the cavity field~\cite{nat404}.
Cavity QED systems have been widely used in many fields, such as
cavity cooling~\cite{jpb38, pra64, nat428, prlZippilli}, atomic dynamics detection
~\cite{ol21} or atomic quantum phase probing~\cite{np3, prl98}.

For the system of an ultra-cold atom in a strong coupling cavity,
the condition of large atomic detuning is often satisfied, which
enables to neglect the influence of the atomic spontaneous emission.
However, when investigating the long time evolution or the steady
state property of the system, spontaneous emission can have notable
effect on the atomic spatial coherence and can not be neglected any
more. The recoil by spontaneously emitted photons in random
directions destroys the atomic spatial coherence and interference
fringes in momentum space may not be observed experimentally.
Moreover, when coherently pumped by a laser field, the photons in
the cavity grows rapidly and the cavity field has great fluctuation.
The approximation of taking the lowest vibrational state in the
Wannier expansion is no longer valid~\cite{oc273, epjd46}. Thus a
fully quantum mechanical model has to be implemented to describe the
cavity QED system and the Monte-Carlo wave function (MCWF) method is
commonly used to simulate the time evolution of such a
system~\cite{smqo1, prl68, pra46, epjd44}.

In this paper the effect of spontaneous emission on the atomic
coherence property in the cavity is studied with a fully quantum
mechanical model. By comparing the time evolution of the atomic
momentum distribution with and without atomic spontaneous emission,
we find that the influence of the atomic spontaneous emission can
not be neglected in evaluating the steady state properties and is
responsible for the lost of spatial coherence. Furthermore, the
dependence of the atomic spatial coherence property on the cavity
parameters is studied. The pumping strength rather than the cavity
decay rate is the dominating factor affecting the atomic locality.

\begin{figure}[tbp]
 \begin{center}
  \includegraphics[width=6cm]{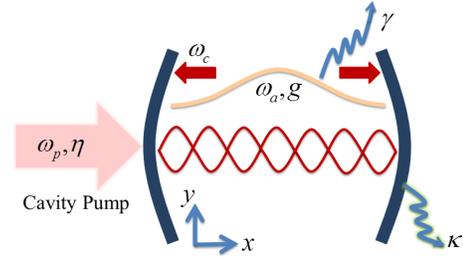}
 \end{center}
\caption{The cavity pump scheme. A two-level atom with transition
frequency $\omega_{\rm a}$ is coupled to a cavity with resonance
frequency $\omega_{\rm c}$, which is coherently pumped by a laser
with frequency $\omega_{\rm p}$ and amplitude $\eta$. The coupling
strength between the atom and the cavity is $g$. The cavity decay
rate is $2\kappa$ and the atomic spontaneous emission rate is
$2\gamma$.} \label{fig:cav_pump_scheme}
\end{figure}

\textbf{The model describing the atom-cavity-field interaction--} We
consider a two-level atom with mass $\mu$ and transition frequency
$\omega_{\rm a}$ coupled to a single mode standing-wave cavity with
resonance frequency $\omega_{\rm c}$ and mode function $f(\hat
{\mathbf r})$ (see Fig.~\ref{fig:cav_pump_scheme}). The coupling
strength between the atom and the cavity field is $g$. The cavity is
pumped coherently by a laser with frequency $\omega_{\rm p}$ and
amplitude $\eta$. The photons can either leak out of the cavity from
the end mirrors directly (cavity decay) or be emitted out of the
cavity by the atom (spontaneous emission decay), with decay rate
$2\kappa$ and $2\gamma$, respectively. The time evolution of the
system is governed by the master equation~\cite{cmp48}
\begin{eqnarray}\label{eq:lindblad_equation}
\dot \rho = \frac{1}{i \hbar} [H, \rho] + {\mathcal L} \rho.
\end{eqnarray}
Using the rotating-wave and electric-dipole approximations, the
Hamiltonian can be depicted in the frame rotating with $\omega_{\rm
p}$ as~\cite{orta, ieee51}
\begin{eqnarray}\label{eq:jc_hamiltonian}
H &=& - \hbar \Delta_{{\rm c}} \hat a^\dag \hat a - i \hbar \eta
\left( \hat a - \hat a ^\dag \right) + \frac{\hat {\mathbf p}^2}{2
\mu} \nonumber\\&-&  \hbar \Delta_{{\rm a}} \hat \sigma_ + \hat
\sigma_- -i \hbar g f(\hat {\mathbf r}) \left( \hat \sigma_+ \hat a
- \hat \sigma_- \hat a^\dag \right),
\end{eqnarray}
where the terms on the right describe by order, the cavity field,
the pumping of the cavity, the atomic motion, the atomic internal
energy and the atom-field coupling. $\Delta_{\rm c}=\omega_{\rm
p}-\omega_{\rm c}$ and $\Delta_{\rm a}=\omega_{\rm p}-\omega_{\rm
a}$ are the cavity and atomic detunings from the frequency of the
pumping laser. $\hat a$ and $\hat a^\dag$ are the annihilation and
creation operators of the cavity field. $\hat \sigma_+$ and $\hat
\sigma_-$ are the raising and lowering operators of the atom. The
Liouvillean is given by~\cite{smqo1}
\begin{eqnarray}\label{eq:liouvillean}
{\mathcal L}\rho &=& \gamma \left( 2\int {\rm d}^2 {\mathbf u}
N({\mathbf u}) \hat \sigma_- {\rm e}^{-i k_{\rm a}{\mathbf u}\hat
{\mathbf r}} \rho {\rm e}^{i k_{\rm a}{\mathbf u}\hat {\mathbf r}}
\hat \sigma_+\right.\nonumber\\ &-& \left. \left[\hat \sigma_+ \hat
\sigma_-, \rho \right]_+\right) + \kappa \left( 2\hat a \rho \hat
a^\dag - \left[ \hat a^\dag \hat a,\rho \right]_+ \right),
\end{eqnarray}
with ${\mathbf u}$ the direction vector of the spontaneously emitted
photons and $N({\mathbf u})$ the directional distribution for the
atomic spontaneous emission, which is considered as an isotropic one
for simplicity. $k_{\rm a}=\omega_{\rm a}/c$ is the wave number
corresponding to the atomic transition. The first term on the right
of Eq.~(\ref{eq:liouvillean}) describes the spontaneous emission
together with the atomic momentum recoil and the second term the
cavity decay.

In our model the atomic motion is restricted along the cavity axis
($x$ direction in Fig.~\ref{fig:cav_pump_scheme}). The cavity mode
function is approximated by a sine mode $f(\hat {\mathbf r})=f(\hat
x)=\sin \left( K \hat x \right)$, with $K$ the wave number of the
cavity field. The recoil of the atom by spontaneously emitted
photons is projected onto the cavity axis. $k_{\rm a}$ can be well
approximated by $K$ since the detuning between the atomic transition
frequency and the cavity resonance frequency is much smaller than
$\omega_{\rm a}$ and $\omega_{\rm c}$. The recoil frequency of the
atom by absorbing or emitting either a photon from either the cavity
field or the pump field is then presented as $\omega_{\rm r}=\hbar
K^2/(2\mu)$. Typical values of $\omega_{\rm r}/(2\pi)$ for
$^{133}$Cs and $^{87}$Rb are $2.0663$~kHz and 3.7710~kHz,
respectively.

In the case of far-off-resonance pumping, the large atomic detuning
leads to low atomic saturation, and we can adiabatically eliminate
the upper atomic level. The lowering operator of the atom is then
presented as~\cite{epjd46, ao}
\begin{eqnarray}
\hat \sigma_- \approx \frac{g f(\hat x) \hat a}{i \Delta_{\rm a} -
\gamma},
\end{eqnarray}
and $\hat \sigma_+ = \hat \sigma_-^\dag$. Inserting these
expressions into Eqs.~({\ref{eq:jc_hamiltonian}) and
Eq.~(\ref{eq:liouvillean}), we can obtain the effective Hamiltonian
\begin{eqnarray}\label{eq:h_eff}
H_{{\rm eff}} = - \hbar \Delta_{\rm c} \hat a^\dag \hat a - i \hbar
\eta \left( \hat a - \hat a^\dag \right) + \frac{\hat p^2}{2 \mu} +
\hbar U_0 f^2(\hat x)\hat a^\dag \hat a,
\end{eqnarray}
and the effective Liouvillean
\begin{eqnarray}\label{eq:l_eff}
{\mathcal L}_{{\rm eff}} \rho &=& \Gamma_0 \left( 2\sum_u N(u)
f(\hat x) \hat
a {\rm e}^{-iKu\hat x} \rho {\rm e}^{iKu\hat x}\right.\nonumber\\
&-& \left.\left[ f^2(\hat x)\hat a^\dag \hat a, \rho
\right]_+\right) + \kappa \left( 2\hat a \rho \hat a^\dag - \left[
\hat a^\dag \hat a,\rho \right]_+ \right),
\end{eqnarray}
with $U_0=g^2\Delta_{\rm a}/\left( \Delta_{\rm a}^2 + \gamma^2
\right)$ the effective atom-field coupling strength and
$2\Gamma_0=2g^2\gamma/\left( \Delta_{\rm a}^2 + \gamma^2 \right)$
the effective spontaneous emission rate. $u$ is the projection of
the direction vector of the spontaneously emitted photons on the $x$
axis. The cavity decay can be described by the jump operator $\hat
J_{\rm c}=\sqrt{2\kappa} \hat a$ and the spontaneous emission by the
operator $\hat J_{\rm a}=\sqrt{2\Gamma_0}{\rm e}^{-iKu\hat x}f(\hat
x)\hat a$. The Liouvillean can be further transformed to the
standard form ${\mathcal L} \rho = \sum_m \left( J_m \rho J_m^\dag -
\frac{1}{2}\left[ J_m^\dag J_m,\rho \right]_+ \right)$.

The state vector of the system is given by $\left| \psi
\right\rangle = \sum_{n,k} C_{n,k}(t)\left| n \right\rangle \left| k
\right\rangle$ where $\left| n \right\rangle$ is the $n$th Fock
state of the cavity field and $\left| k \right\rangle$ is the $k$th
atomic momentum state, corresponding to a momentum  $p = k \hbar K$.
As in ~\cite{jpb38}, the integration in Eq.~(\ref{eq:l_eff}) is
reduced to the summation over $u=-1,0,1$. We assume the cavity field
to be in the vacuum state and the atom to be in the zero-momentum
state initially. Because of atomic momentum diffusion in the
periodic potential, very high dimension for describing the momentum
Hilbert space is needed (in our simulation the dimension is taken to
be $2^6$). The Fock basis for the cavity field is truncated up to
the 10th or 20th state. With Monte-Carlo wave function method we can
simulate the time evolution for a stochastic trajectory of the state
vector. According to the ergodic hypothesis, the dynamical process
of the system can be expressed using the time-dependent density
operator $\rho(t)$, which is given approximately by averaging over a
large number of trajectories, and the steady-state property of the
system can be expressed by the steady-state density operator
$\rho_{{\rm ss}}$, which is approximated by averaging over a long
time for one trajectory~\cite{jpb38}.

\begin{figure}[tbp]
 \begin{center}
  \begin{picture}(0,0)
   \put(-5,75){(a)}\quad
  \end{picture}
  \includegraphics[width=4.15cm]{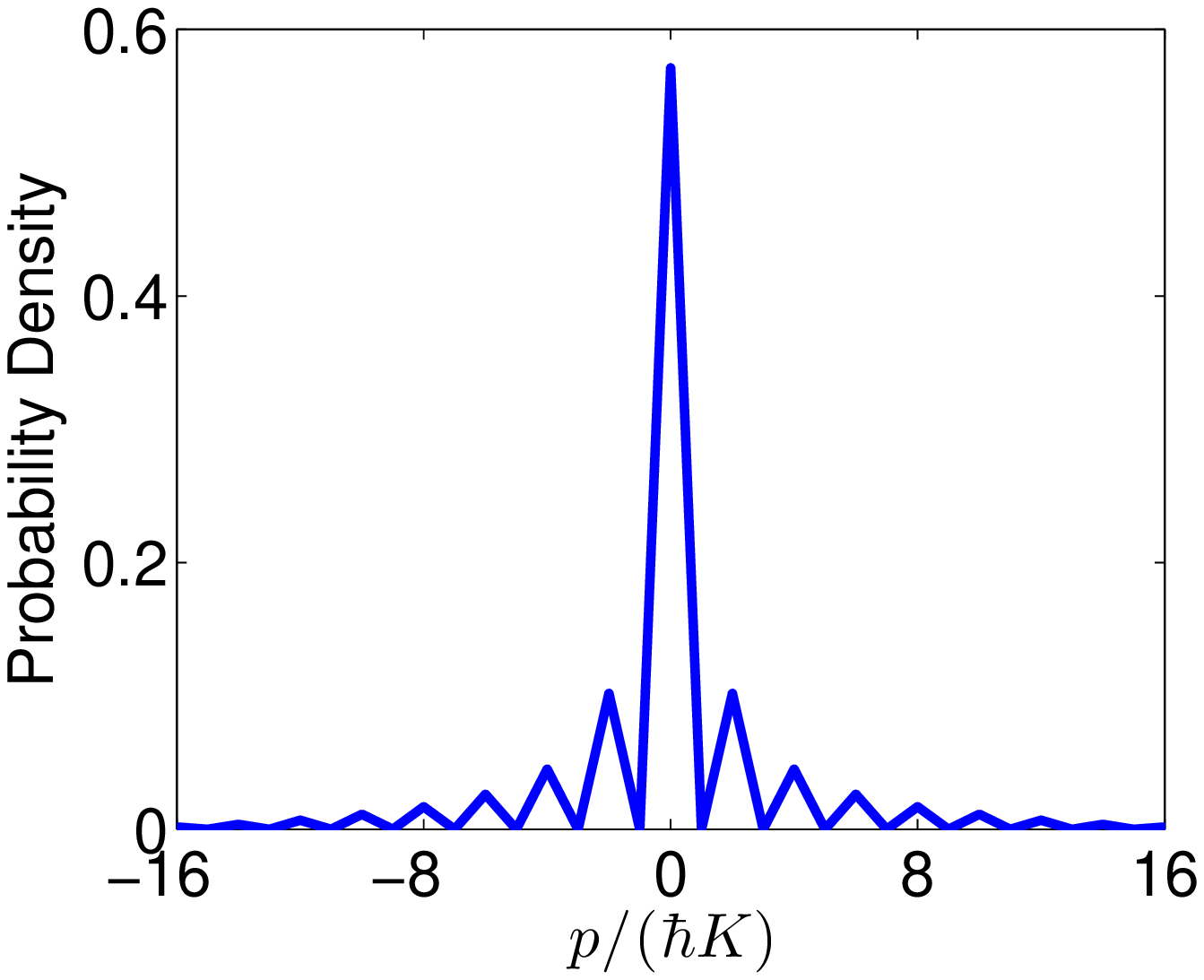}
  \begin{picture}(0,0)
   \put(-5,75){(d)}\quad
  \end{picture}
  \includegraphics[width=4.15cm]{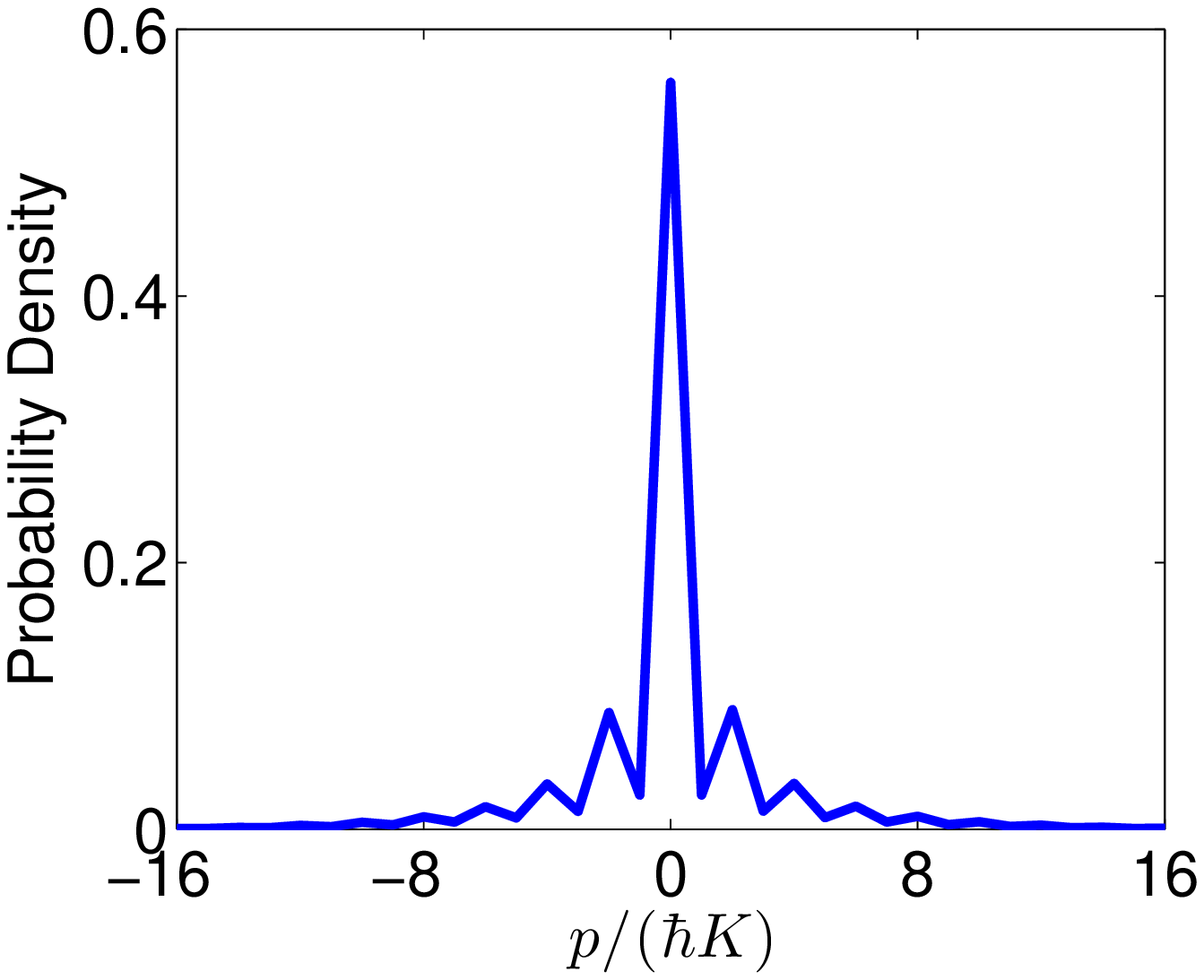}\\
  \begin{picture}(0,0)
   \put(-5,75){(b)}\quad
  \end{picture}
  \includegraphics[width=4.15cm]{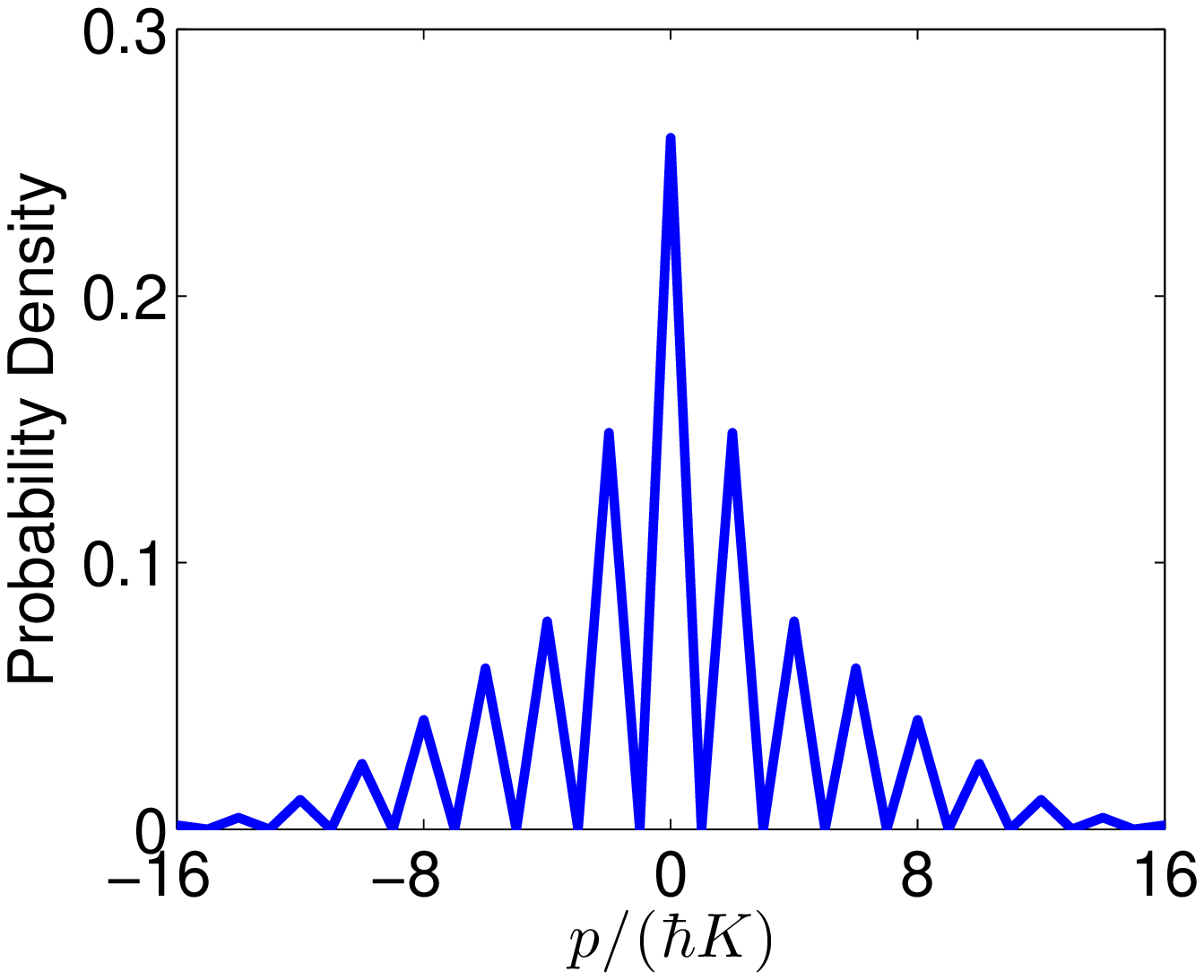}
  \begin{picture}(0,0)
   \put(-5,75){(e)}\quad
  \end{picture}
  \includegraphics[width=4.15cm]{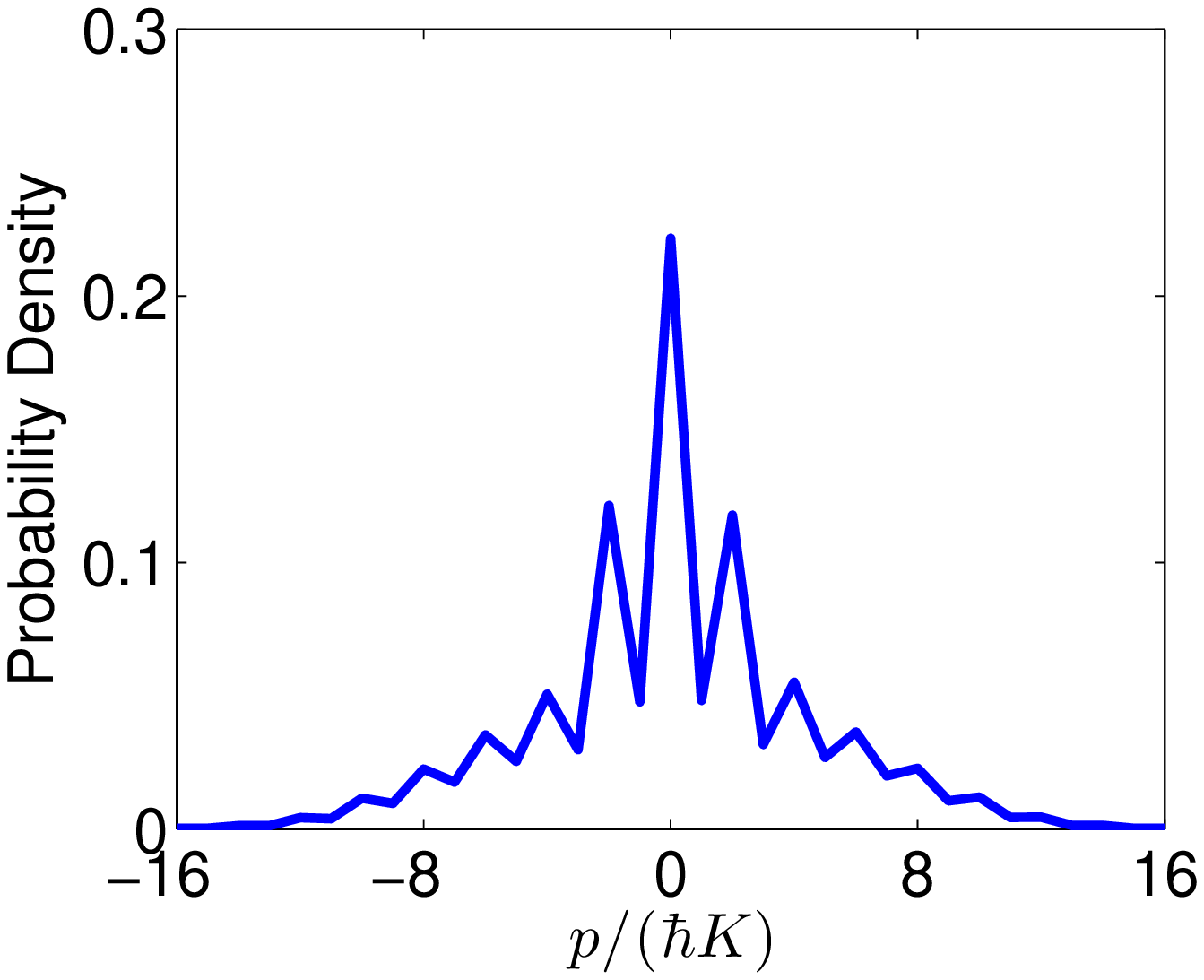}\\
  \begin{picture}(0,0)
   \put(-5,75){(c)}\quad
  \end{picture}
  \includegraphics[width=4.15cm]{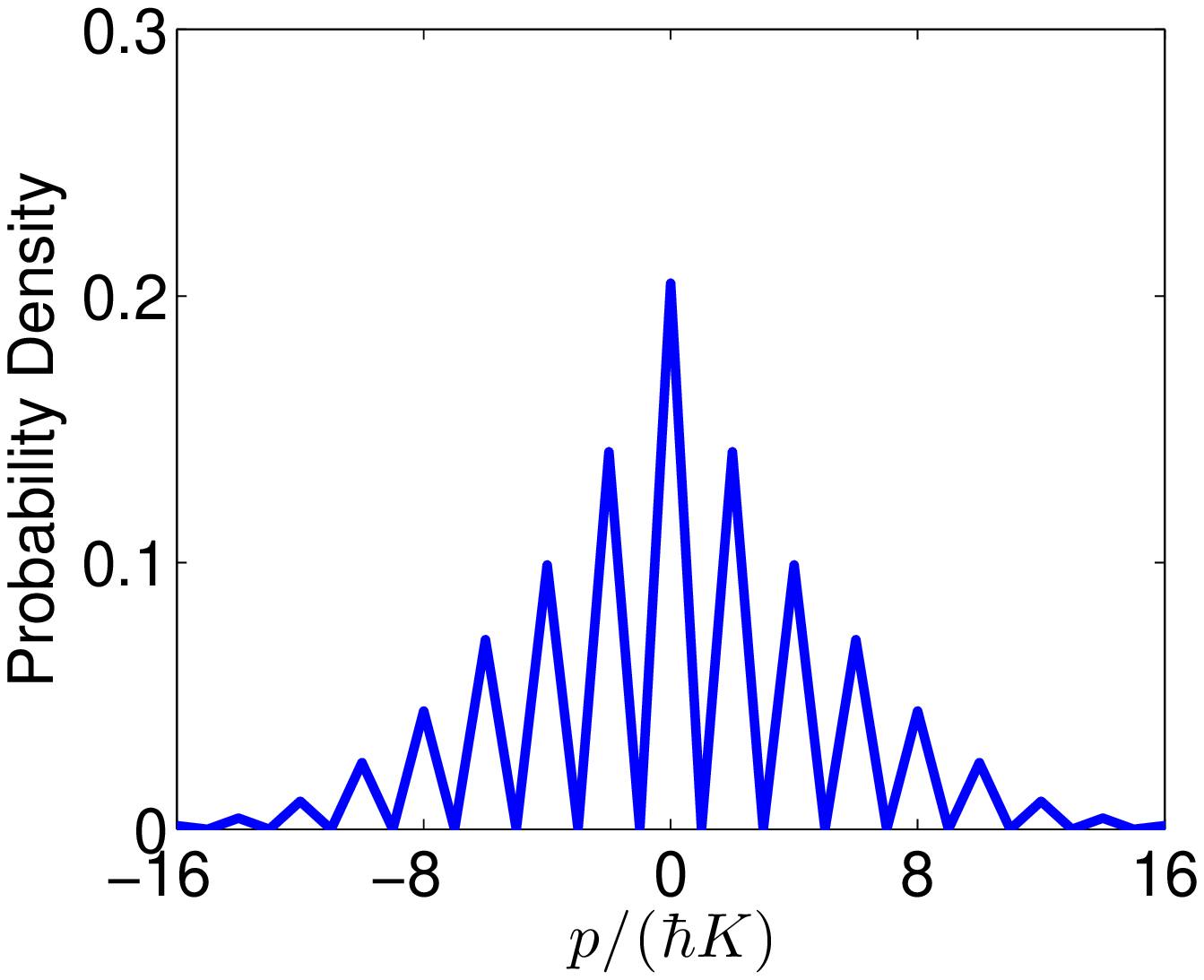}
  \begin{picture}(0,0)
   \put(-5,75){(f)}\quad
  \end{picture}
  \includegraphics[width=4.15cm]{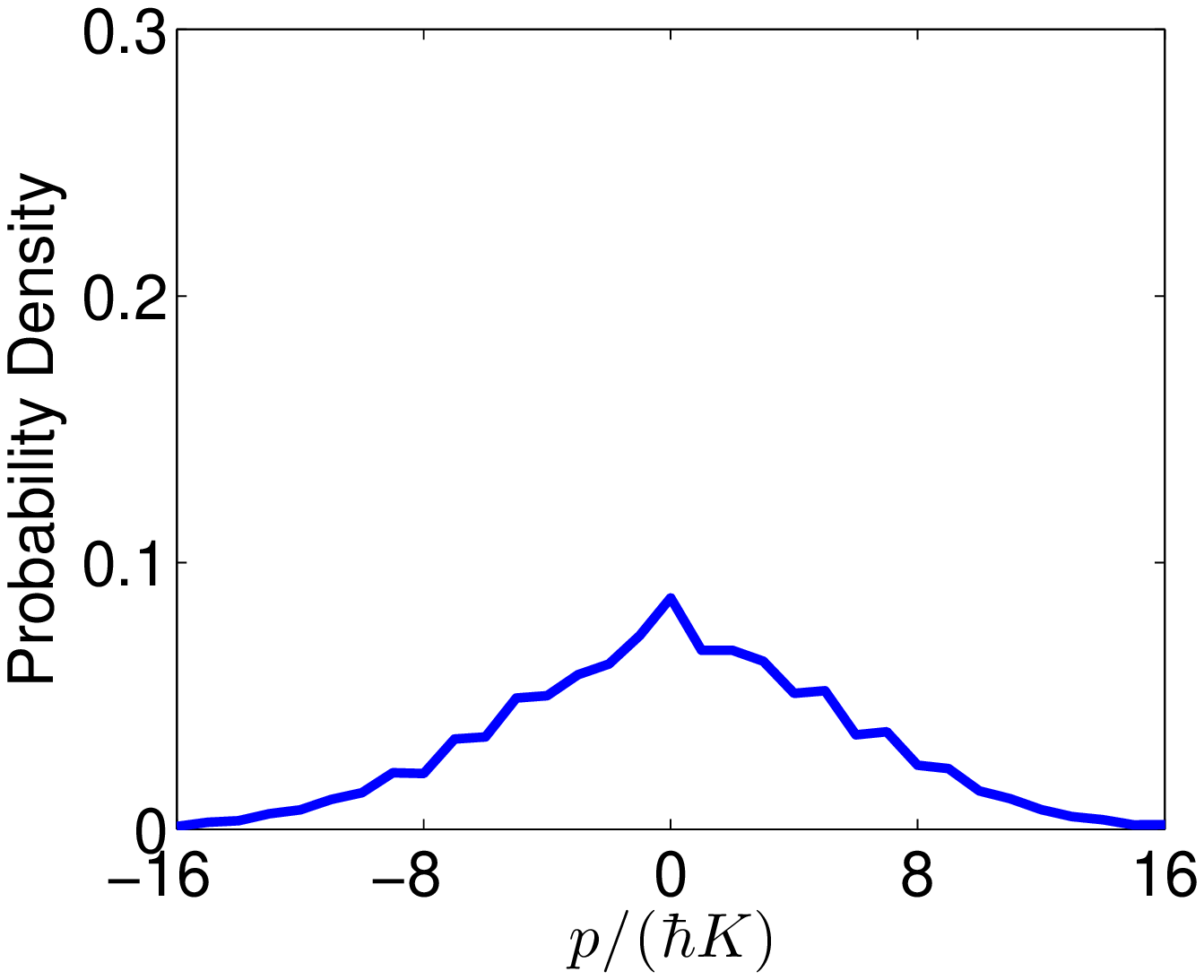}
 \end{center}
\caption{The atomic momentum distribution with $\Gamma_0=0$ (a-c)
and $\Gamma_0=18.75\omega_{\rm r}$ (d-f) for $\omega_{\rm
r}t=0.032$, $0.16$ and $0.72$ from top to bottom. All results are
given after averaging over 200 trajectories. The vertical axis
represents the probability density. $\kappa=31.25\omega_{\rm r}$,
$\eta=62.5\omega_{\rm r}$, $\Delta_{\rm c} = U_0 = -390\omega_{\rm
r}$.} \label{fig:mom_evol_cav_pump}
\end{figure}

\textbf{Dynamics and steady state of the probability density--} In
order to show clearly the effects of the atomic spontaneous
emission, we present results with and without spontaneous emission,
respectively. The time evolution of the atomic momentum
distribution, that is, the diagonal elements of $\rho(t)$, is
plotted in Fig.~\ref{fig:mom_evol_cav_pump}. When the atomic
spontaneous emission is neglected, the interference fringes in
momentum space are formed with peaks at $p = 2m \hbar K(m=0,\pm
1,...)$ along with the establishment of the periodic potential in
the cavity. Compared with the result of an optical lattice potential
in free space~\cite{apb73}, high-order momentum can be enhanced due
to the strong atom-field coupling in the cavity.

When atomic spontaneous emission is considered, the recoil of the
atom in random directions breaks the periodicity of the atomic
spatial distribution. Thus, the spatial coherence of the atomic
distribution is destroyed and the probability density is similar to
a thermal equilibrium distribution. However, in the early stage of
the establishment of the cavity field, because the spontaneous
emission rate is much smaller than the atom-field coupling strength,
the interference fringes can still be observed with lower visibility
as shown by Fig.(2d) and (2e).

\begin{figure}[tbp]
 \begin{center}
  \begin{picture}(0,0)
   \put(-5,75){(a)}\quad
  \end{picture}
  \includegraphics[width=4.15cm]{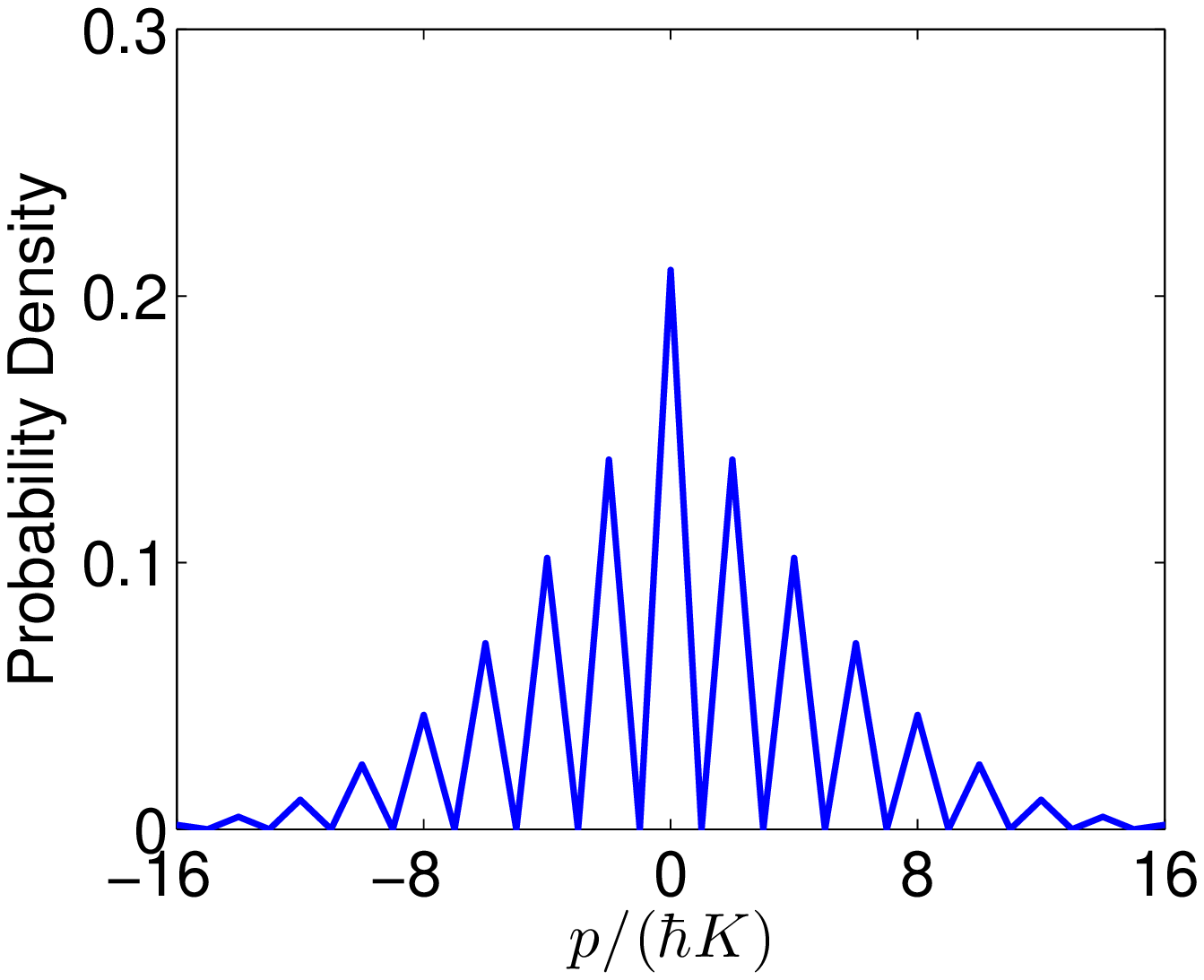}
  \begin{picture}(0,0)
   \put(-5,75){(b)}\quad
  \end{picture}
  \includegraphics[width=4.15cm]{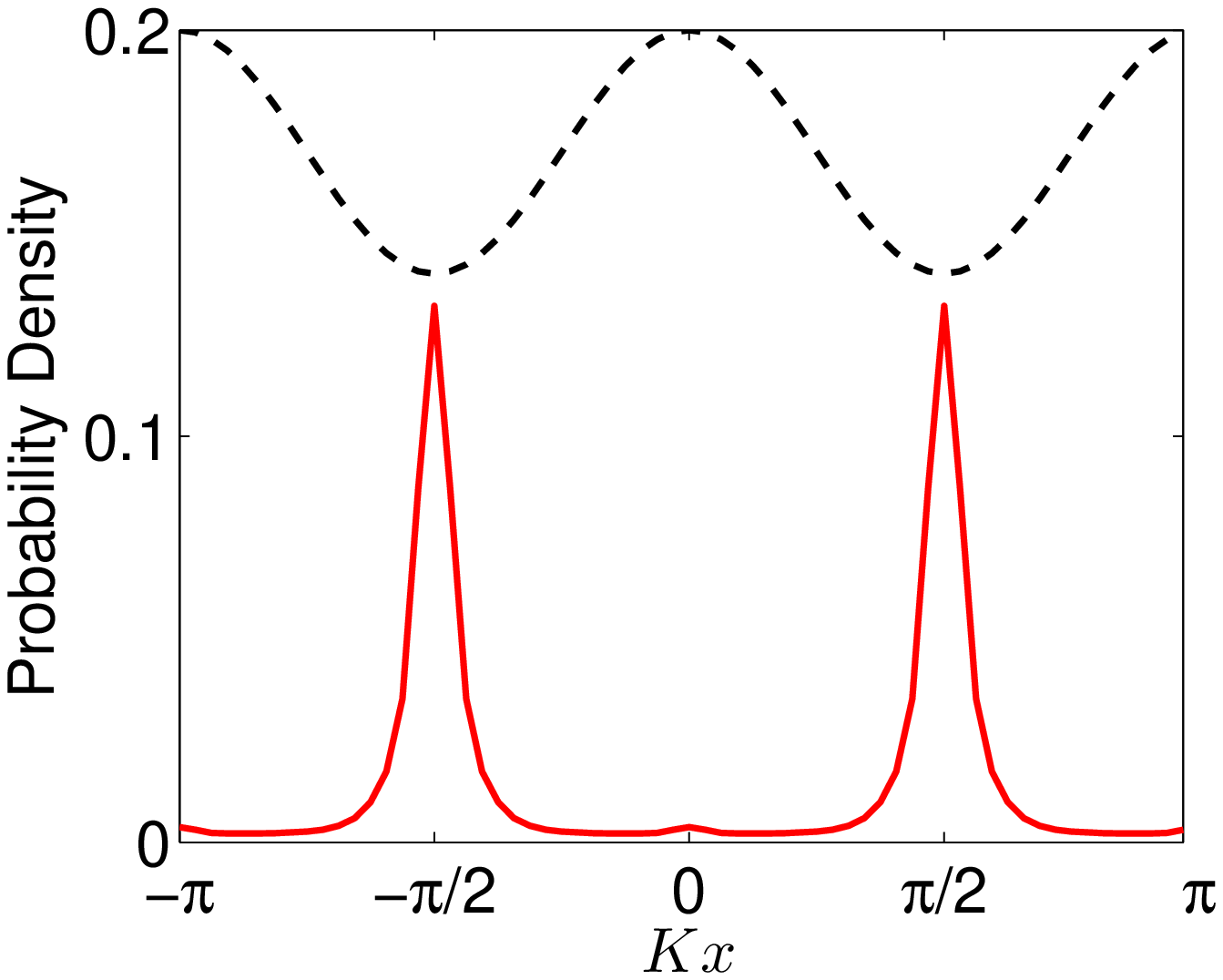}\\
  \begin{picture}(0,0)
   \put(-5,75){(c)}\quad
  \end{picture}
  \includegraphics[width=4.15cm]{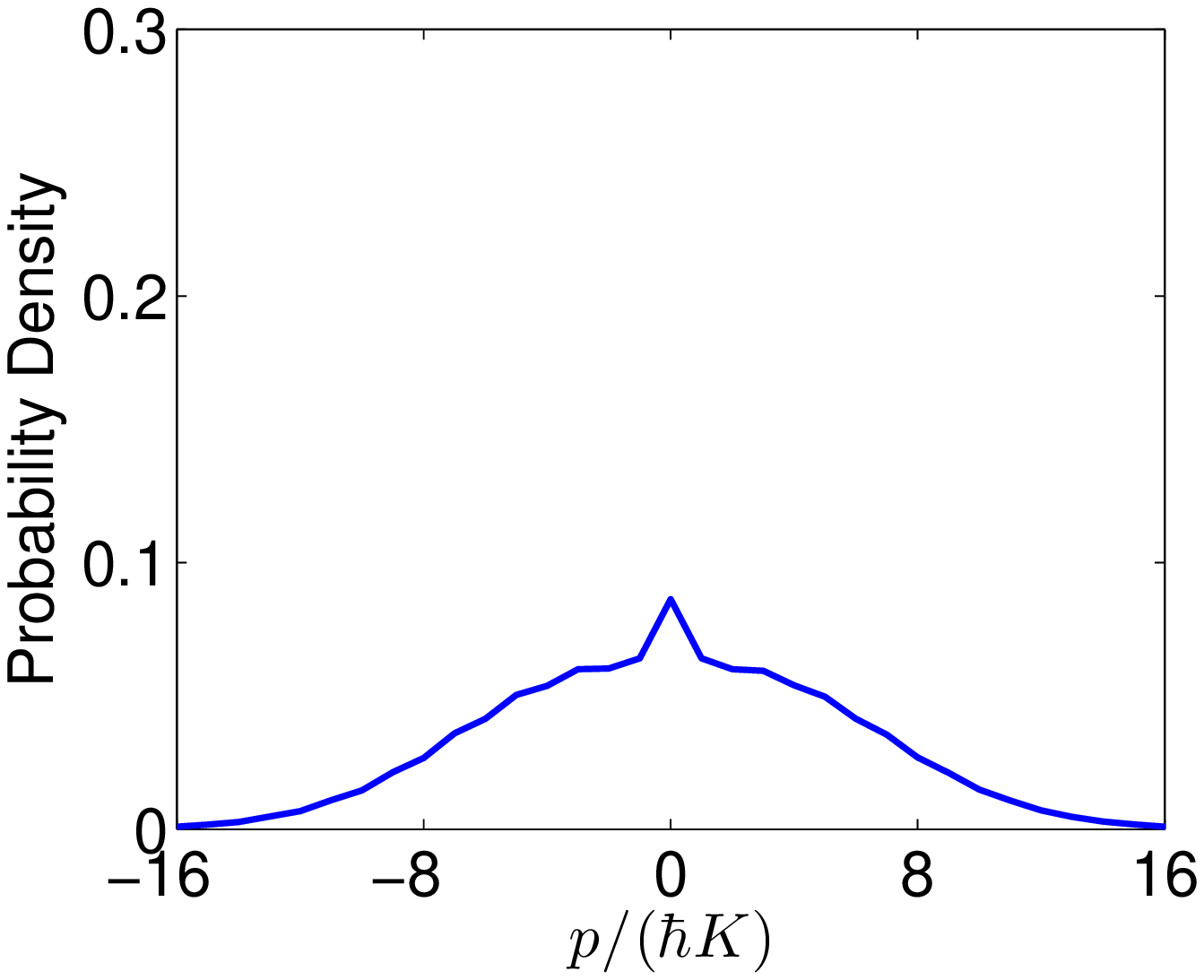}
  \begin{picture}(0,0)
   \put(-5,75){(d)}\quad
  \end{picture}
  \includegraphics[width=4.15cm]{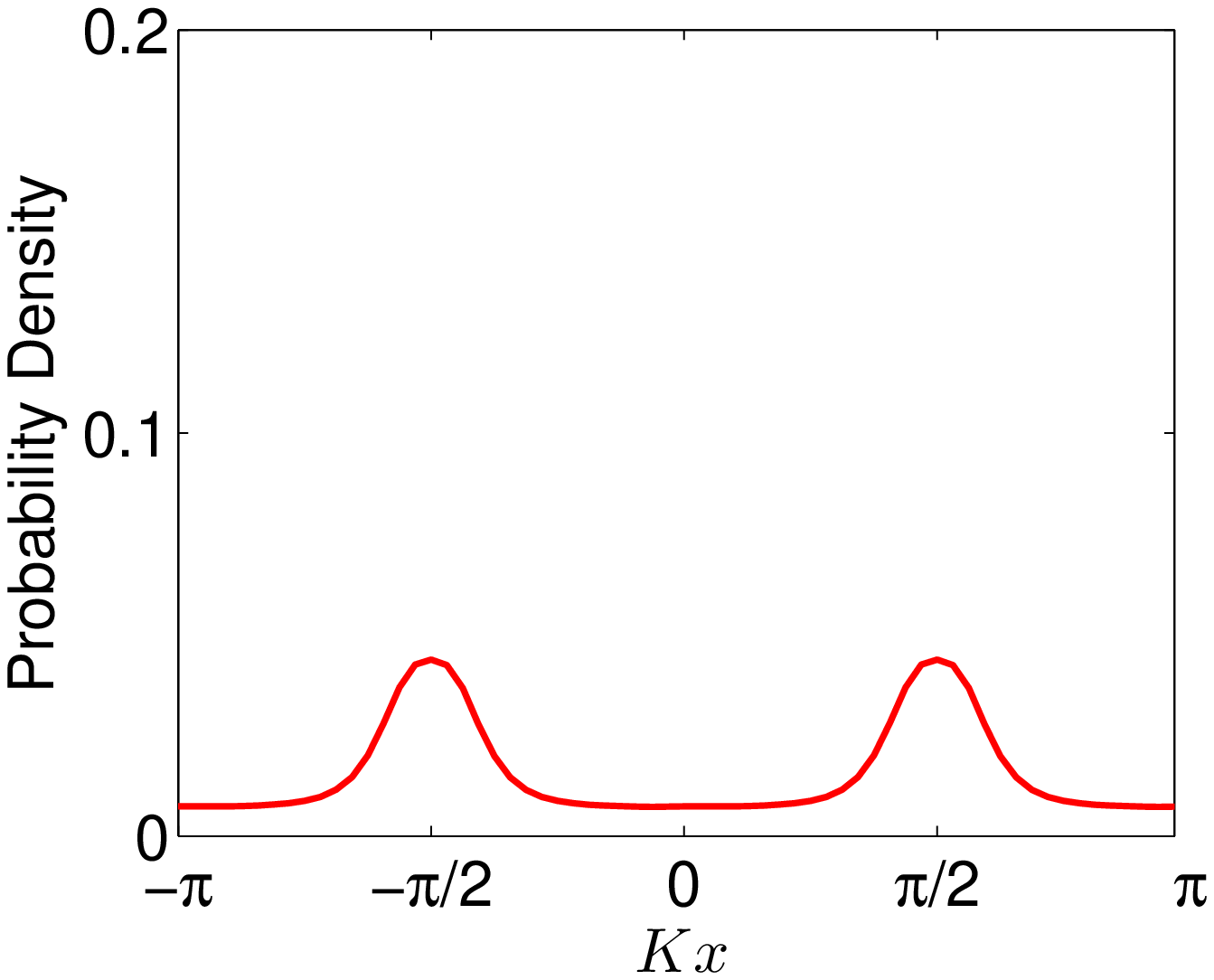}
 \end{center}
\caption{The probability density versus the atomic momentum and
spatial distribution of the steady state. (a) and (b) represent the
atomic momentum and spatial distribution for $\Gamma_0=0$,
while (c) and (d) show the same curves for $\Gamma_0=18.75\omega_{\rm r}$.
The dashed line in (b) shows the potential. $\kappa=31.25\omega_r$,
$\eta=62.5\omega_r$, $\Delta_c = U_0 = -390\omega_r$.}
\label{fig:ss_dist}
\end{figure}
The spatial and momentum distribution for the steady state is given
in Fig.~\ref{fig:ss_dist}. The peaks of the probability density are
localized in the center of the lattice sites. When $\Gamma_0=0$, the
coherence between different sites results in interference fringes in
momentum space (see Fig.~(\ref{fig:ss_dist}a)). Nevertheless, with
notable atomic spontaneous emission which may destroy the coherence
among the sites, no fringes can be observed and the heating effect
is depicted as shown in Fig.~(\ref{fig:ss_dist}c). Besides, with the
same $\kappa$ and $\eta$ as well as non-zero $\Gamma_0$, the total
decay rate is larger and the average photon number is smaller, thus
the peaks in Fig.~(\ref{fig:ss_dist}c) are smaller than in
Fig.~(\ref{fig:ss_dist}a).

\begin{figure}[tbp]
 \begin{center}
  \begin{picture}(0,0)
   \put(-5,75){(a)}\quad
  \end{picture}
  \includegraphics[width=4.15cm]{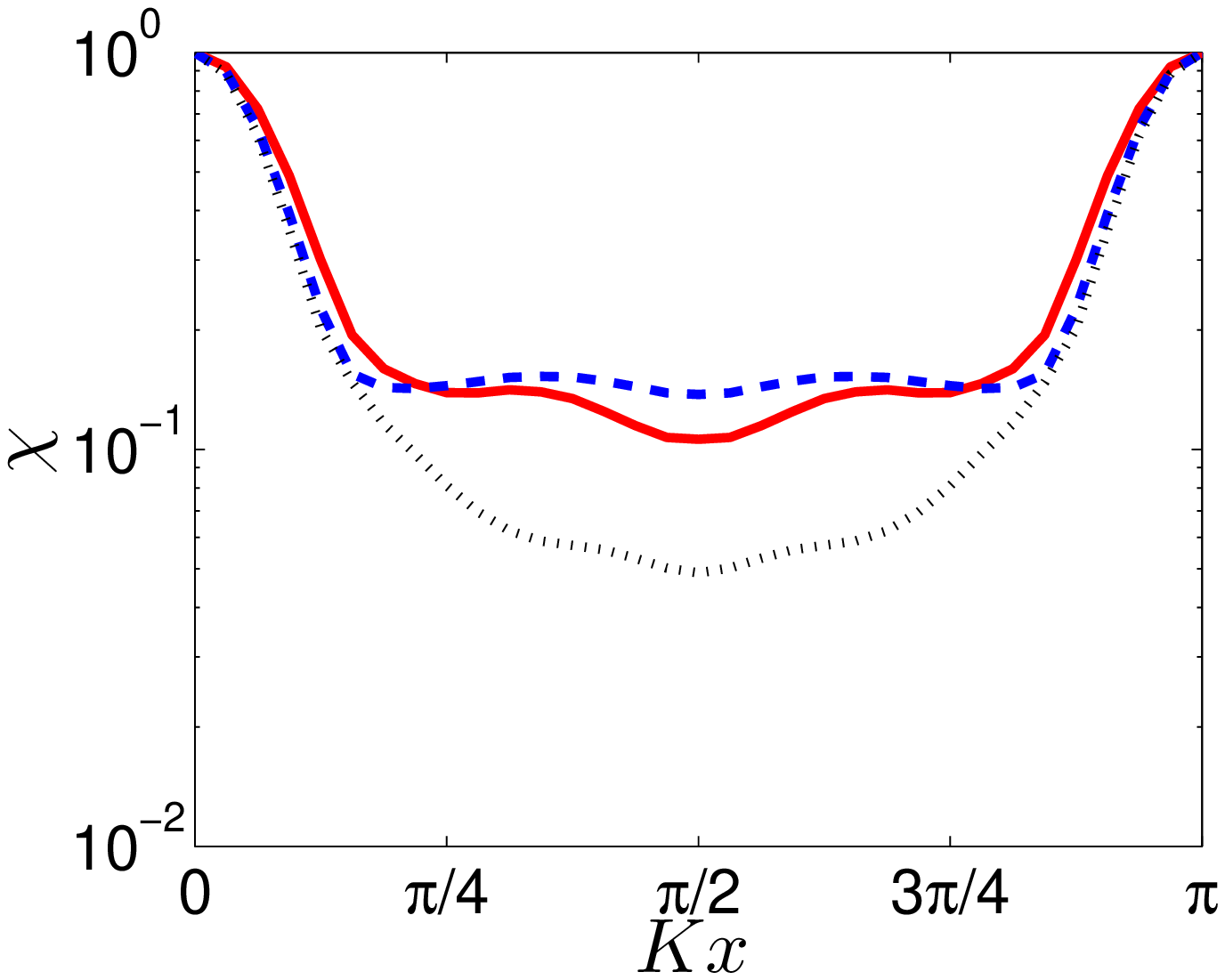}
  \begin{picture}(0,0)
   \put(-5,75){(b)}\quad
  \end{picture}
  \includegraphics[width=4.15cm]{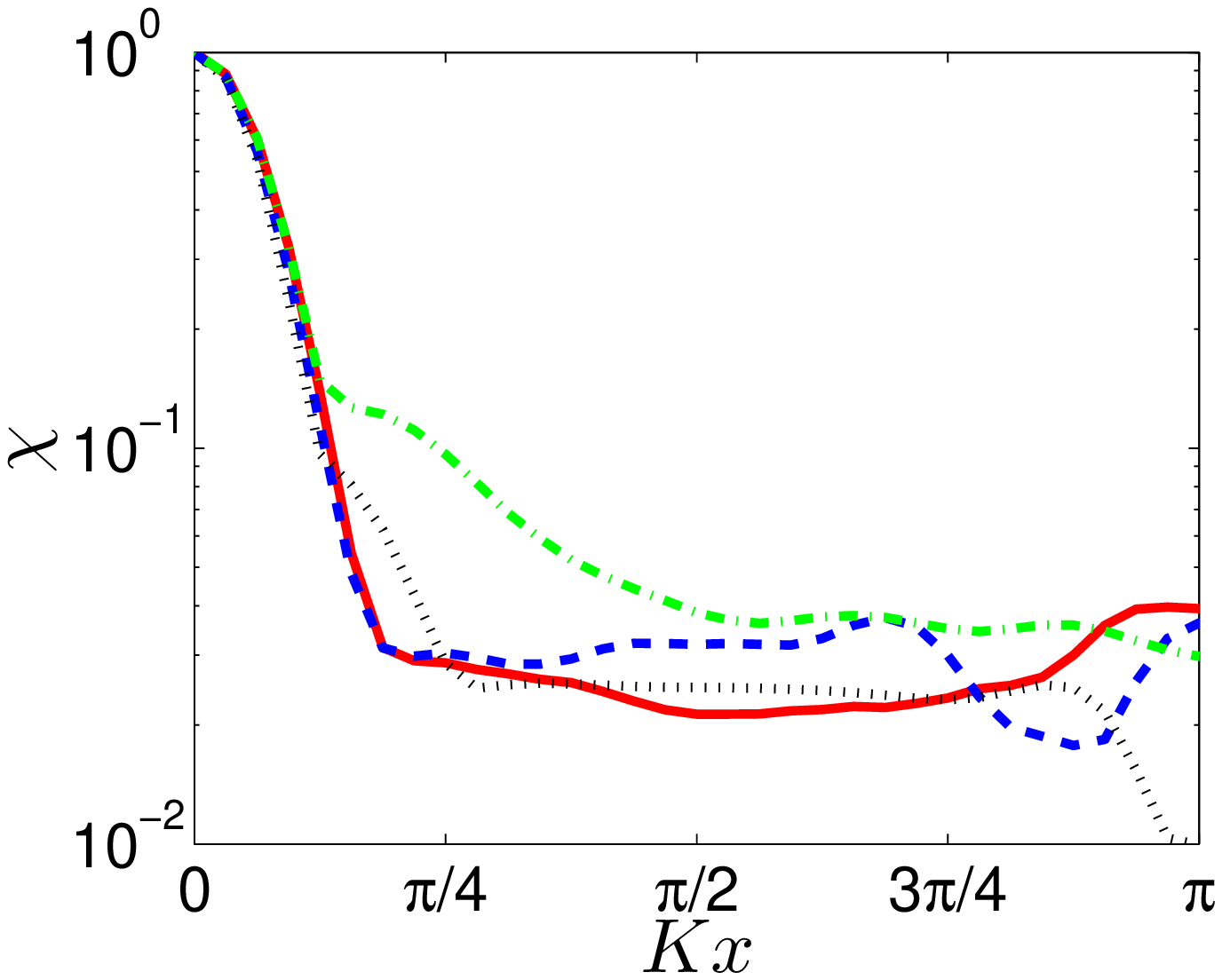}
 \end{center}
\caption{Atomic spatial coherence functions with $\Gamma_0=0$ for
(a) and $\Gamma_0=18.75\omega_{\rm r}$ for (b). The curves indicate
different cavity decay rate and pumping amplitude $(\kappa,
\eta)=(0,31.25\omega_{\rm r})$, $(31.25\omega_{\rm r},
31.25\omega_{\rm r})$, $(62.5\omega_{\rm r}, 31.25\omega_{\rm r})$,
and $(31.25\omega_{\rm r}, 62.5\omega_{\rm r})$ (dash-dotted, solid,
dashed, and dotted lines, respectively).
$U_0=\Delta_c=-390\omega_r$.} \label{fig:coh_func_cav_pump}
\end{figure}

\textbf{Influence of the system parameters on the atomic coherence
property--} The atomic spatial coherence property can be measured by
the coherence function $\chi(x)$~\cite{jpb38}
\begin{eqnarray}\label{eq:corr_func}
\chi(x)=\int {\rm d}\left( K\xi \right) \left| \rho_{\rm
a}(\xi,\xi+x) \right|,
\end{eqnarray}
where $\rho_{\rm a}(x_1,x_2) = \left\langle x_1 \right| \left(
\sum_n \left\langle n \right| \rho \left| n \right\rangle \right)
\left| x_2 \right\rangle$ is the reduced density matrix describing
the atomic spatial distribution. The coherence between neighboring
sites is given by $\chi(x = \lambda_{\rm c}/2 =\pi / K)$. The
coherence function for different parameters is depicted in
Fig.~\ref{fig:coh_func_cav_pump}. When the spontaneous emission is
neglected, the coherence between neighboring sites is conserved
($\chi(\pi / K)=1$). However, when considering the influence of
spontaneous emission, the coherence between neighboring sites
vanishes ($\chi(\pi / K) \ll 1$).

We can perform an integration for the coherence function to get the
spatial coherence degree
\begin{eqnarray}
C = \frac{1}{\pi} \int_0^{\pi} {\rm d}\left( Kx \right) \chi(x),
\end{eqnarray}
which reflects the average coherence property over a period of the
atomic spatial distribution.

\begin{figure}[tbp]
 \begin{center}
  \begin{picture}(0,0)
   \put(-5,75){(a)}\quad
  \end{picture}
  \includegraphics[width=4.15cm]{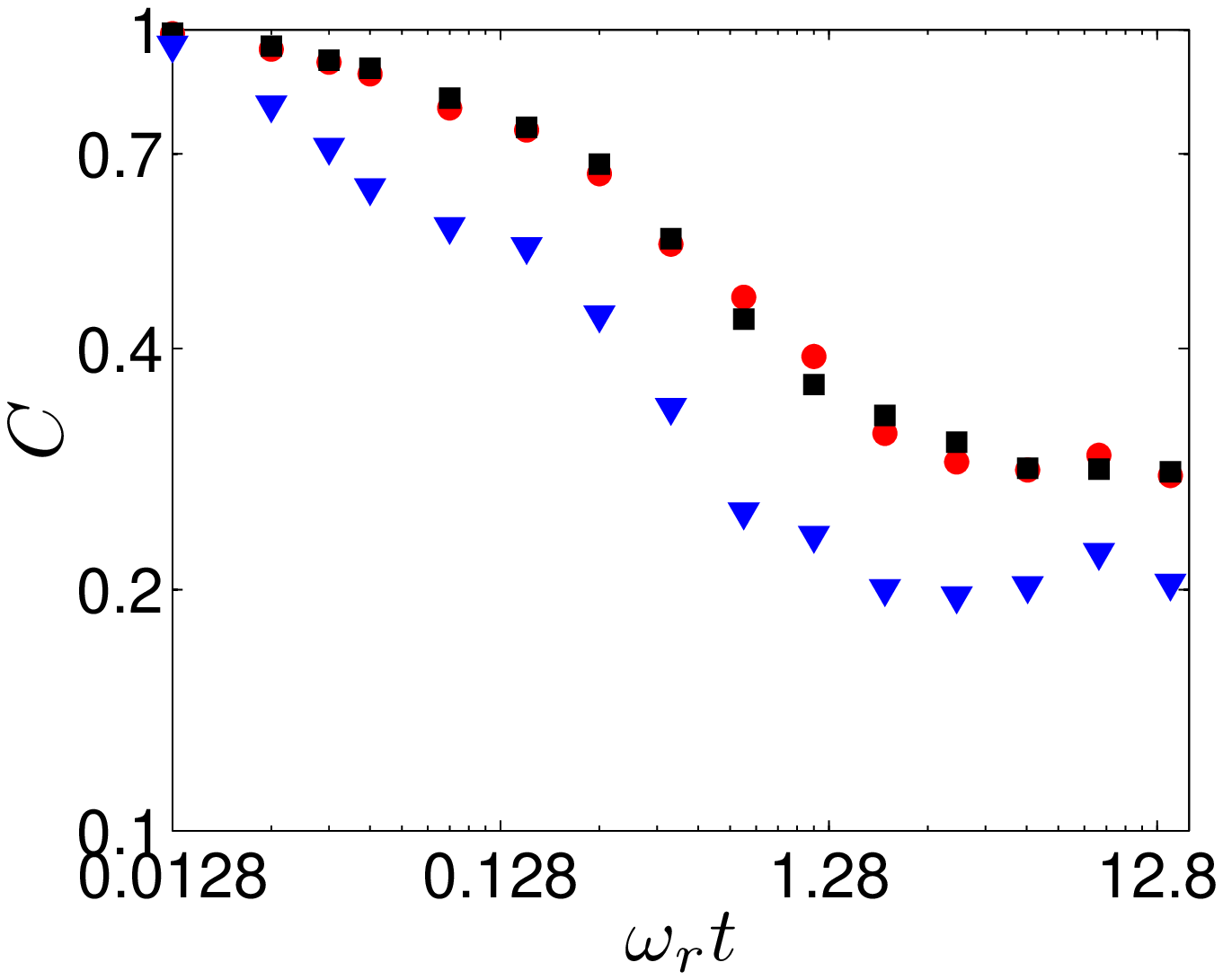}
  \begin{picture}(0,0)
   \put(-5,75){(b)}\quad
  \end{picture}
  \includegraphics[width=4.15cm]{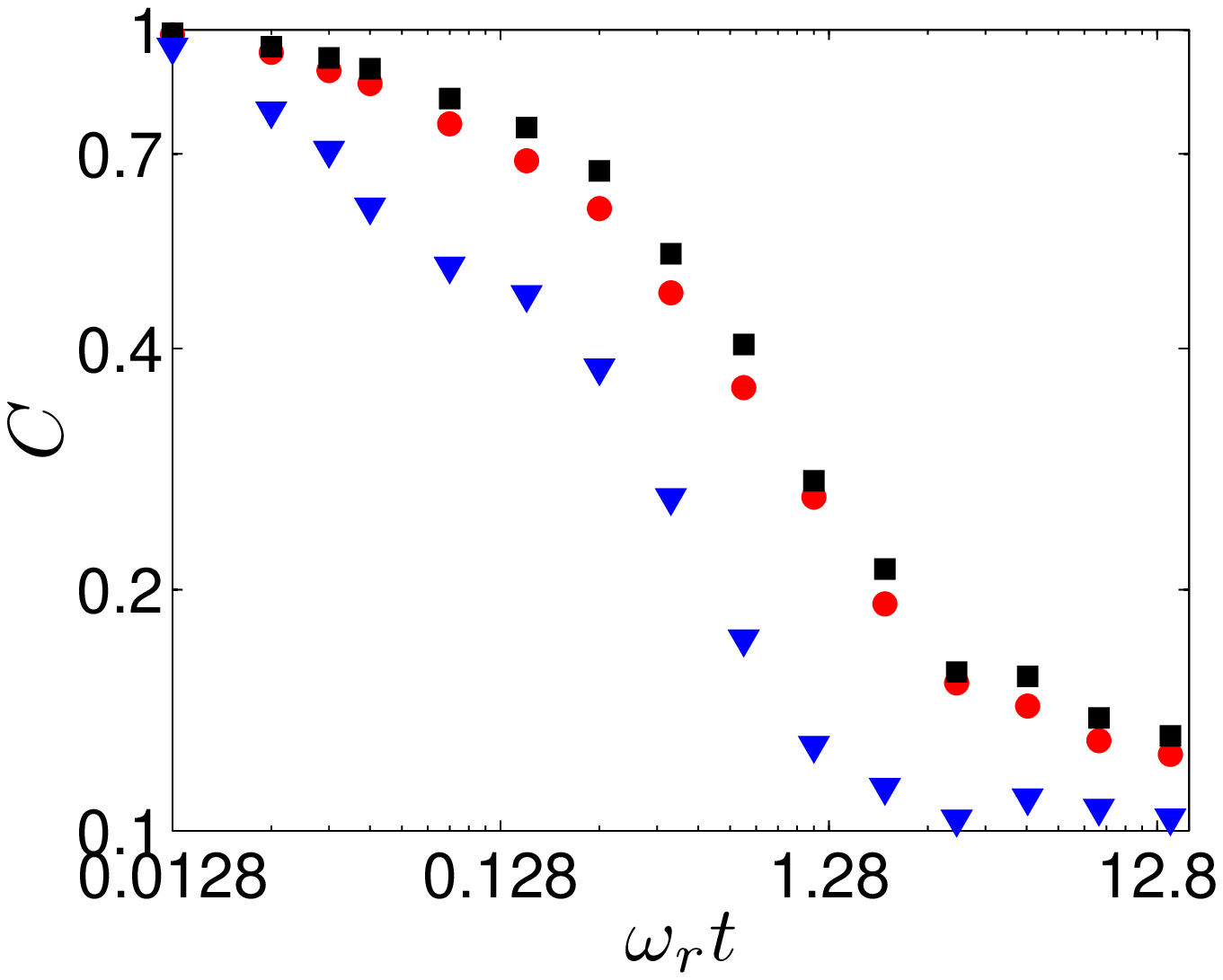}
 \end{center}
\caption{The time evolution of the atomic coherence degree with
$\Gamma_0=0$ for (s) and $\Gamma_0=18.75\omega_{\rm r}$ for (b). All
results are given after averaging over 200 trajectories. The curves
indicates different cavity decay rate and pumping amplitude
$(\kappa, \eta)=(0,31.25\omega_{\rm r})$, $(31.25\omega_{\rm r},
31.25\omega_{\rm r})$, $(62.5\omega_{\rm r}, 31.25\omega_{\rm r})$,
and $(31.25\omega_{\rm r}, 62.5\omega_{\rm r})$ (diamond, circle,
square, and triangle lines, respectively). $\Delta_c = U_0 =
-390\omega_r$.} \label{fig:c_evol_cav}
\end{figure}

The time evolution of the atomic spatial coherence degree is shown
in Fig.~\ref{fig:c_evol_cav}. With the establishment of the lattice
in the cavity, the peaks for the probability density are localized
in the center of the sites, and the nonuniform distribution leads to
the decrease of the atomic spatial coherence. When the effect of
spontaneous emission is considered, the phase of the atomic wave
function at different sites is changed randomly due to the recoil,
which may further decrease the coherence degree.

Now we investigate the influence of the cavity decay rate $\kappa$
and the pumping amplitude $\eta$. From Eq.~(\ref{eq:h_eff}) and
(\ref{eq:l_eff}) we know that the pumping amplitude and the cavity
decay do not influence the atomic spatial or momentum distribution
directly, but influence the atom through the coupling term $\hbar
U_0 f^2(\hat x) \hat a^\dag \hat a$. With large cavity decay, the
cavity field adiabatically follows the atomic motion, and from the
Heisenberg equation of $\hat a^\dag$ and $\hat a$ we have
\begin{eqnarray}
\hat a^\dag \hat a=\frac{\eta^2}{\kappa^2 + \left[ \Delta_{\rm
c}-U_0 \sin^2(K\hat x) \right]^2}.
\end{eqnarray}
Thus, even for the resonance situation of $\Delta_{\rm c}=U_0$, the
spatial spread of the atomic probability density still causes a
shift of the cavity resonance frequency, which can be much larger
than $\kappa$. Consequently, the cavity decay rate may have little
influence on the atomic spatial distribution and the atomic
coherence property. Fig.~\ref{fig:coh_func_cav_pump} and
\ref{fig:c_evol_cav} show that the atomic coherence property does
not depend much on $\kappa$ at fixed pumping strength $\eta$.
However, for larger pumping strength $\eta$, the photon number in the cavity is larger,
resulting in deeper potential for the optical lattice in the cavity.
The peaks of the atomic spatial distribution become sharper, resulting in smaller
coherence length. Therefore, the atomic spatial coherence degree
decreases.

\textbf{Discussion and conclusion--} The dynamics and steady state
property for the atomic momentum and spatial distribution as well as
the atomic spatial coherence have been investigated with MCWF
method. By comparing the results of situations with and without
spontaneous emission, we find that the atomic spontaneous emission
is dominant during the decoherence process. Besides, due to the
atomic spatial spread of the probability distribution, the pumping
strength is found to have greater influence on the photon number in
the cavity and then the atomic locality than the cavity decay rate.
The spontaneous emission should be suppressed in experiment when the
long time evolution of the atomic spatial coherence property is
investigated. In fact, by normalizing the atomic wave-function to
the particle number $N$ and modifying the effective coupling
strength $U_0$ in Eq.~(\ref{eq:h_eff}) to the collective one $NU_0$,
this model can also be used to investigate the coupling between a
non-interacting BEC and the quantized cavity field. With the method
of absorption imaging and coherent measurement technology of cavity
QED ~\cite{qo}, the results may be directly observed and tested by
experiments.

We are grateful to M. Z. Wu and T. Vogt for critical reading our
manuscript. This work is partially supported by the state Key
Development Program for Basic Research of China (No.2005CB724503,
2006CB921402 and 2006CB921401), NSFC(10874008 and 10934010), SRF
for ROCS and SEM.


\begin{references}
\bibitem{prl95} C. Maschler, and H. Ritsch, Phys. Rev. Lett. {\bf 95}, 260401 (2005).

\bibitem{nat404} P. Pinkse, et. al, Nature {\bf 404}, 365 (2002).

\bibitem{pra64} V. Vuleti\'c, H. W. Chan, and A. T. Black, Phys. Rev. A {\bf 64}, 033405 (2001).

\bibitem{nat428} P. Maunz, et. al, Nature {\bf 428}, 50 (2004).

\bibitem{prlZippilli}  S. Zippilli,  G. Morigi, Phys. Rev. Lett. {\bf 95}, 143001 (2005).

\bibitem{jpb38} A. Vukics, J. Janszky. and P. Domokos, J. Phys. B {\bf 38}, 1453 (2005).

\bibitem{ol21} H. Mabuchi, et. al, Opt. Lett. {\bf 21} 1393 (1996).

\bibitem{np3} I. B. Mekhov, C. Maschler, and H. Ritsch, Nature Physics {\bf 3}, 319 (2007).

\bibitem{prl98} I. B. Mekhov, C. Maschler, and H. Ritsch, Phys. Rev. Lett. {\bf 98}, 100402 (2007).

\bibitem{oc273} C. Maschler, et. al, Opt. Commun. {\bf 273}, 446 (2007).

\bibitem{epjd46} C. Maschler, I. B. Mekhov, and H. Ritsch, Eur. Phys. J. D {\bf 46}, 545 (2008).

\bibitem{prl98-2} S. Slama, et. al, Phys. Rev. Lett. {\bf 98}, 053603 (2007).

\bibitem{pra80} Xiaoji Zhou, Phys. Rev. A {\bf 80}, 023818 (2009).

\bibitem{smqo1} H. J. Carmichael, Statistical Methods in Quantum Optics: Master Equations and Fokker-Planck Equations, Springer, (1999).

\bibitem{prl68} J. Dalibard, Y. Castin, and K. M\/{o}lmer, Phys. Rev. Lett. {\bf 68}, 580 (1992).

\bibitem{pra46} C. W. Gardiner, A. S. Parkins, and P. Zoller, Phys. Rev. A {\bf 46}, 4363 (1992).

\bibitem{epjd44} A. Vukics, and H. Ritsch, Eur. Phys. J. D {\bf 44}, 585 (2007).

\bibitem{cmp48} G. Lindblad, Commun. Math. Phys. {\bf 48}, 119 (1976).

\bibitem{orta} L. Allen, and J. H. Eberly, Optical Resonance and Two-Level Atoms, Dover Publications, Inc. New York, (1975).

\bibitem{ieee51} E. T. Jaynes, and F. W. Cummings, Proc. IEEE {\bf 51}, 89 (1963).

\bibitem{ao} P. Meystre, Atom Optics, Springer (2001).

\bibitem{apb73} M. Greiner, \emph{et al.} Appl. Phys. B {\bf 73}, 769 (2001).

\bibitem{qo} D. F. Walls, and G. J. Milburn, Quantum Optics. Springer-Verlag, (1995).


\end{references}
\end{document}